\documentclass[aps,pra,superscriptaddress,reprint,showpacs,longbibliography,amsmath,amssymb]{revtex4-1}
\usepackage[utf8]{inputenc}
\usepackage[T1]{fontenc}
\usepackage{microtype}
\usepackage{graphicx}
\usepackage{mathbbol}
\usepackage[colorlinks,linkcolor=blue,citecolor=blue,urlcolor=blue]{hyperref}

\def\equationautorefname~#1\null{Eq.~(#1)\null}

\newcommand{\braket}[1]{\ensuremath{\left\langle{#1}\right\rangle}}
\newcommand{\bra}[1]{\langle #1 |}
\newcommand{\ket}[1]{| #1 \rangle}

\begin{document}

\title{Exploiting vibrational strong coupling to make an optical parametric oscillator out of a Raman laser}

\author{Javier del Pino}
\affiliation{Departamento de Física Teórica de la Materia Condensada and Condensed Matter Physics Center (IFIMAC), Universidad Autónoma de Madrid, E-28049 Madrid, Spain}
\author{Francisco J. Garcia-Vidal}
\email{fj.garcia@uam.es}
\affiliation{Departamento de Física Teórica de la Materia Condensada and Condensed Matter Physics Center (IFIMAC), Universidad Autónoma de Madrid, E-28049 Madrid, Spain}
\affiliation{Donostia International Physics Center (DIPC), E-20018
  Donostia/San Sebastián, Spain}
\author{Johannes Feist}
\email{johannes.feist@uam.es}
\affiliation{Departamento de Física Teórica de la Materia Condensada and Condensed Matter Physics Center (IFIMAC), Universidad Autónoma de Madrid, E-28049 Madrid, Spain}

\begin{abstract}
  When the collective coupling of the rovibrational states in organic
  molecules and confined electromagnetic modes is sufficiently strong,
  the system enters into vibrational strong coupling, leading to the
  formation of hybrid light-matter quasiparticles. In this work we
  demonstrate theoretically how this hybridization in combination with
  stimulated Raman scattering can be utilized to widen the
  capabilities of Raman laser devices. We explore the conditions under
  which the lasing threshold can be diminished and the system can be
  transformed into an optical parametric oscillator. Finally, we show
  how the dramatic reduction of the many final molecular states into
  two collective excitations can be used to create an all-optical
  switch with output in the mid-infrared.
\end{abstract}


\maketitle

When the coherent interaction between a confined light mode and
vibrational matter excitations becomes faster than the relevant
decoherence processes, the system can enter into vibrational strong
coupling (VSC)~\cite{Shalabney2015,George2015a,Long2015,DelPino2015,
  Simpkins2015,Muallem2016}. The fundamental excitations of the two
systems then become inextricably linked and can be described as hybrid
light-matter quasiparticles, so-called vibro-polaritons, that combine
the properties of both ingredients. In particular, the use of
vibrational modes that are both IR- and Raman-active allows to probe
vibro-polaritons through Raman scattering mediated by their material
component~\cite{Shalabney2015a,DelPino2015b,Strashko2016}.

On the other hand, while the cross sections for Raman scattering are
typically small, the process can become highly efficient under strong
driving if the scattered Stokes photons accumulate sufficiently to
lead to stimulated Raman scattering (SRS)~\cite{Penzkofer1979}. The
effective energy conversion from input to output beam can then be
exploited to fabricate a highly tunable Raman laser. Raman lasers have
been realized using a variety of nonlinear media and configurations,
such as under pulsed operation in optical fibers~\cite{Stolen1972},
nonlinear crystals~\cite{Pask2003}, gases~\cite{Benabid2002}, or
silicon~\cite{Boyraz2004}, as well as under continuous-wave operation
in silicon~\cite{Rong2005,Rong2007}, silica~\cite{Kippenberg2004} and
molecular hydrogen~\cite{Brasseur1998}. Since the threshold powers for
these systems are typically large, they suffer from detrimental
effects such as Kerr nonlinearities, four-wave mixing, and heat
deposition~\cite{Pask2003}.

In this Letter, we propose and theoretically demonstrate that the
hybrid light-matter nature of vibro-polaritons can be exploited to
obtain photon emission from the vibrationally excited final states of
a Raman laser. A single-output Raman laser device then becomes
analogous to an \emph{optical parametric oscillator}
(OPO)~\cite{Boyd2008} with output beams both in the visible and in the
mid-IR, relevant for many spectroscopic
applications~\cite{Sorokina2003}. In addition to obtaining two
coherent beams with a stable phase relation (and possibly
nonclassical correlations~\cite{Walls1983,Reid1988,Xiao1987})
spanning very different frequency regions, this approach has the
further advantage of effectively getting rid of the energy deposited
into material vibrations; instead of being dissipated as heat, this
energy is emitted in the form of photons. Finally, we show that the
coexistence of the upper and lower polariton modes with very similar
properties can be exploited to produce an all-optical
switch~\cite{Gibbs1985,Dawes2005}. Here, one (gate) pump beam can be
used to switch Raman lasing of a second (signal) pump beam.

\begin{figure}
  \includegraphics[width=\linewidth]{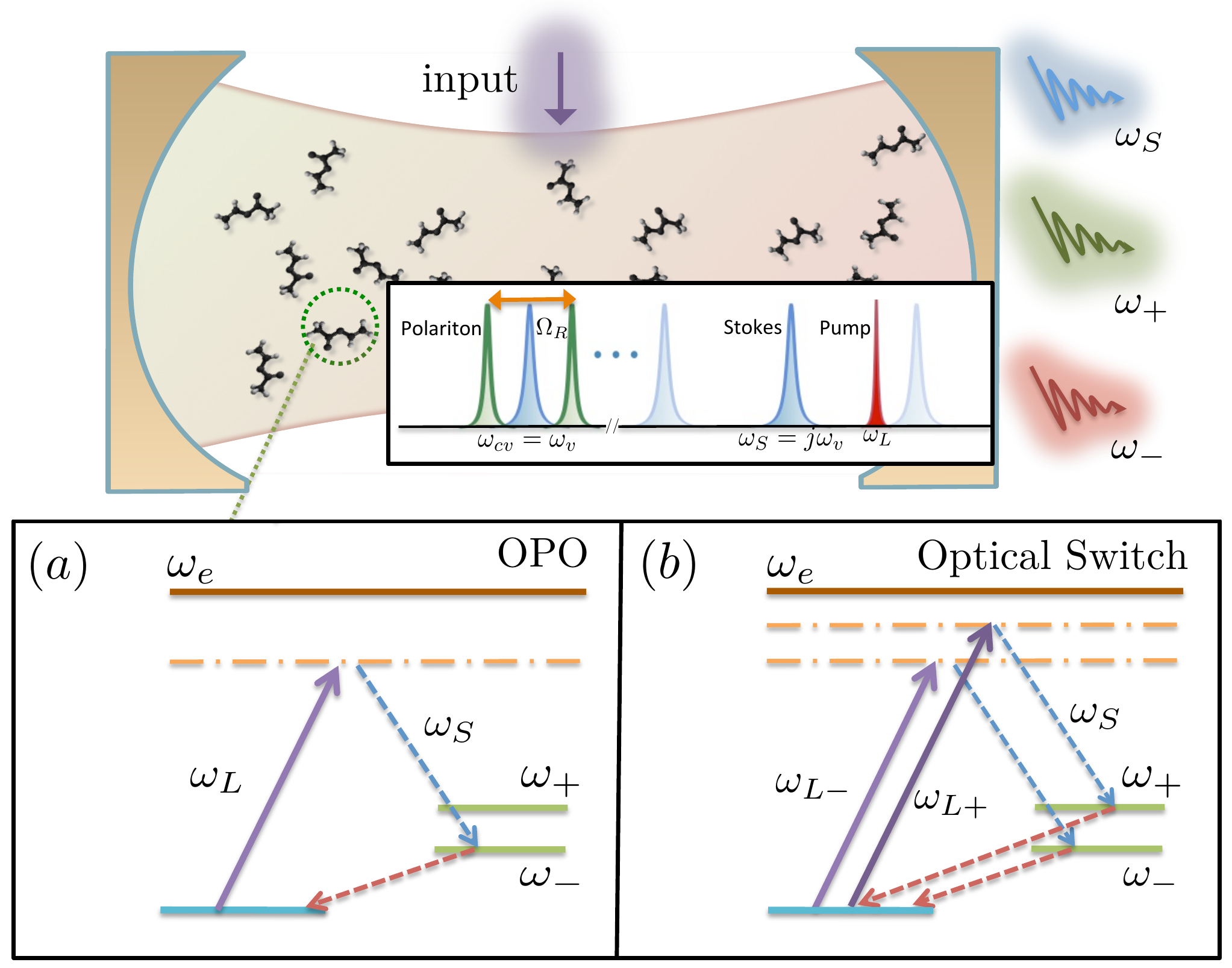}
  \caption{Upper panel: sketch of the system to convert a Raman laser
    into an OPO through vibrational strong coupling (see main
    text). The input fields (purple arrows) can be chosen to achieve
    (a) OPO operation with a single pump frequency $\omega_L$, or (b)
    an all-optical switch with two pump fields
    $\omega_{L\pm}$.}\label{fig:setup}
\end{figure}

The system we consider (sketched in \autoref{fig:setup}) consists of
a material with a vibrational transition that is both IR- and
Raman-active, placed inside a resonator (e.g., a microcavity). The
resonator supports at least two confined modes, a mid-IR mode used to
achieve VSC with the vibrational transition, and an optical mode used
to accumulate the scattered Stokes photons. We model the material as a
set of $N$ noninteracting three-level quantum emitters, formed by the
ground state $\ket{g}$ (energy $\omega_g\equiv0$), the first excited
vibrational mode $\ket{v}$ (energy $\omega_v$), and an electronically
excited state $\ket{e}$ (energy $\omega_e$)~\cite{DelPino2015b}.
While this model can naturally represent organic molecules (as used in
current experiments achieving VSC), we note that it can also be used
to treat systems such as the nonlinear crystals utilized in existing
Raman lasers. The IR-active ground-vibrational transition is
resonantly coupled to the mid-IR cavity mode at frequency $\omega_c$,
with annihilation operator $\hat{a}_c$. The Hamiltonian describing the
vibrational excitations and their strong coupling to the mid-IR mode
within the rotating wave approximation (RWA) is given by (setting
$\hbar = 1$ here and in the following)
\begin{equation}\label{eq:system_H}
  \hat{H}_s = \omega_c\hat{a}_c^{\dagger}\hat{a}_c + 
              \sum_{i=1}^N\left[ \omega_v \hat{\sigma}_{vv}^{(i)} + 
              \left(g \hat{a}_c^{\dagger}\hat{\sigma}_{gv}^{(i)} +
              \mathrm{H.c.}\right)\right].
\end{equation}
Here, $\hat{\sigma}_{ab}^{(i)} = \ket{a^{(i)}}\bra{b^{(i)}}$ denotes
the transition operator between the states $\ket{b}$ and $\ket{a}$ of
the $i$th molecule, while the light-matter interaction strength is
measured by $g$, which depends on the single-photon electric field
strength of the mid-IR cavity mode and the change of the molecular
dipole moment under displacement from the equilibrium position.

Assuming zero detuning ($\omega_c=\omega_v$) for simplicity, the
eigenstates of $\hat{H}_s$ are formed by \emph{i)} two
vibro-polaritons,
$\ket{\pm}=\frac{1}{\sqrt{2}}(\hat{a}_c^{\dagger}\ket{G}\pm\ket{B})$,
symmetric and antisymmetric hybridizations of the cavity mode with the
collective \emph{bright state} of the molecular vibrations,
$\ket{B}=\frac{1}{\sqrt{N}}\sum_{i=1}^N\ket{v^{(i)}}$. Here, $\ket{G}$
denotes the global ground state. The polaritons have eigenfrequencies
$\omega_\pm=\omega_v\pm g\sqrt{N}$, separated by the Rabi splitting
$\Omega_R= 2g\sqrt{N}$. The other eigenstates are \emph{ii)} $N-1$
so-called \emph{dark states} $\ket{d}$ orthogonal to $\ket{B}$ that
have eigenfrequencies $\omega_v$ and no electromagnetic component.

We first treat the dynamics of the system under external driving of a
single pump mode at frequency $\omega_L$ (not resonant with any cavity
mode), see \autoref{fig:setup}(a). The full Hamiltonian then contains
$\hat{H}_s$ as well as the electronic excitations of the molecules,
the pump field (which we quantize in order to be able to describe
depletion~\cite{Carmichael2008}), the cavity mode in the optical
(frequency $\omega_S$), and the interactions between the
molecular transitions and the optical modes, leading to
\begin{multline}\label{eq:fullH}
  \hat{H} = \hat{H}_s + \omega_S \hat{n}_S + \omega_{L} \hat{n}_L \\
  + \sum_{i=1}^N\left[\omega_e \hat{\sigma}_{ee}^{(i)} + \left(g_{S}\hat{a}_S\hat{\sigma}_{ve}^{(i)\dagger} + g_{L}\hat{a}_{L} \hat{\sigma}_{ge}^{(i)\dagger}+
    \mathrm{H.c.}\right) \right],
\end{multline}
where $\hat{n}_L=\hat{a}_{L}^{\dagger}\hat{a}_{L}$ and
$\hat{n}_S=\hat{a}_S^{\dagger}\hat{a}_S$ are the photon number
operators for the pump laser and confined cavity mode, which are
coupled (within the RWA) to the ground-excited and excited-vibrational
transitions in the molecules, respectively. We assume continuous-wave
driving of the pump mode,
\begin{equation}
  \hat{H}_d=\Phi_{\mathrm{in}}\sqrt{\kappa_L} (\hat{a}_{L}e^{-i\omega_{L} t} + \hat{a}_{L}^{\dagger}e^{i\omega_{L} t}),\label{eq:H_d}
\end{equation}
where $\Phi_{\mathrm{in}}$ parametrizes the driving strength. The
results derived below are also valid under time-dependent driving as
long as the pump amplitude $\Phi_{\mathrm{in}}$ varies more slowly
than the time required to reach the steady state.

When the driving laser is far off-resonant to the electronic transition
such that the hierarchy condition $\omega_e\gg\omega_L\gg\omega_v$ is
satisfied, we can adiabatically eliminate the electronically excited
states from the problem \cite{Brion2007,Reiter2012}. If the laser
frequency is chosen such that Raman scattering to one of the
polaritonic modes is resonant with cavity mode $S$, i.e.,
$\omega_L=\omega_S+\omega_p$, with $p\in\{+,-\}$, scattering to the
other polaritonic mode can then be neglected under a second RWA. This
gives the following effective Hamiltonian (for details see the
supplemental material~\cite{supplemental}):
\begin{multline}
  \hat{H}_{\mathrm{eff}} \simeq
  \omega_L\hat{n}_L + \omega_S\hat{n}_S + \omega_{p}\hat{\sigma}_{pp}
  - g_{\mathrm{eff}}^p
  \left(\hat{a}_{L}\hat{a}_{S}^{\dagger}\hat{\sigma}_{Gp}^{\dagger} +
    \mathrm{H.c.}\right),\label{eq:OPO_eff}
\end{multline}
where the effective coupling occurs between laser photons and pairs of
Stokes photons and polaritons, while the dark modes are not
excited. The coupling strength is given by
\begin{equation}\label{eq:int_OPO}
  g_{\mathrm{eff}}^{p} = \sqrt{\frac{N}{2}}
  \frac{g_{S}g_{L}}{\omega_{e}-(\omega_{p}+\omega_{S})},
\end{equation}
and is not sensitive to the Rabi splitting of the polaritons. This
agrees with the case of linear Raman scattering, where theory predicts
a redistribution of the scattering cross section of the system without
further enhancement~\cite{DelPino2015b,Strashko2016}. When deriving
$g_{\mathrm{eff}}^{p}$, we have assumed perfect spatial overlap between the
three involved modes $L$, $S$, and $p$; inclusion of the spatial
profile would lead to the renormalization
$g_{\mathrm{eff}}^\eta \rightarrow g_{\mathrm{eff}}^\eta \mathcal{S}$,
with $\mathcal{S}$ the overlap integral.

The trilinear interaction in \autoref{eq:OPO_eff} is analogous to a
nondegenerate OPO, converting an input laser beam into two new modes,
the ``signal'' (Stokes beam) and ``idler''
(vibrations)~\cite{Boyd2008}. While this analogy is
well-known~\cite{Penzkofer1979}, it is merely formal for a standard
Raman laser (i.e., in the weak-coupling regime) since most of the
excitation in the vibrationally excited states decays nonradiatively,
such that no idler beam is emitted. In the VSC regime, however, the hybrid
light-matter nature of the polariton imbues them with a photonic
component, leading to efficient outcoupling in the form of
photons. This makes the analogy complete and provides an approach
towards converting a Raman laser into an OPO.

We next discuss the role of losses and dephasing. Within the standard
Lindblad master-equation formalism, the density operator $\hat{\rho}$
evolves according to
\begin{equation}\label{eq:master}
  \partial_t\hat{\rho} = -i[\hat{H}_{\mathrm{eff}},\hat{\rho}] + \kappa_S\mathcal{L}_{\hat{a}_S}[\hat{\rho}]+\kappa_L\mathcal{L}_{\hat{a}_L}[\hat{\rho}]+\tilde{\Gamma}_{\mathrm{vib}}[\hat{\rho}],
\end{equation}
where
$\mathcal{L}_X[\hat{\rho}] = \hat{X}\hat{\rho} \hat{X}^{\dagger} -
\frac{1}{2} \{ \hat{X}^{\dagger}\hat{X}, \hat{\rho}\}$.
The loss rates of the Stokes and quantized laser modes are given by
$\kappa_S$ and $\kappa_L$, respectively. The term
$\tilde{\Gamma}_{\mathrm{vib}}$ summarizes all decoherence mechanisms
affecting the vibrationally excited subspace. Under weak coupling,
these consist of nonradiative decay
($\gamma_v\mathcal{L}_{\hat{\sigma}_{gv}}$) and pure dephasing
($\gamma_{\varphi}\mathcal{L}_{\hat{\sigma}_{vv}}$). In the VSC
regime, the influence of inhomogeneous broadening and dephasing can be
suppressed for large enough Rabi
splitting~\cite{Houdre1996,DelPino2015}, leading to an effective decay
of the polaritons ($\Gamma_\pm\mathcal{L}_{\hat{\sigma}_{g\pm}}$) with
a rate as small as $\Gamma_\pm\approx\frac{\kappa_c+\gamma_v}{2}$,
significantly below the average of the bare-molecule
($\gamma_v+\gamma_\varphi$) and mid-IR cavity ($\kappa_c$) linewidths.

In order to characterize the threshold condition and quantum yield of
the VSC-based OPO described above, we calculate the steady-state
mode populations within the mean-field-approximation, in which all
fields are assumed to be described by coherent amplitudes. In terms of
the slowly-varying amplitudes
$\alpha_L = \braket{\hat{a}_L} e^{i\omega_Lt}$,
$\alpha_S = \braket{\hat{a}_S} e^{i\omega_St}$, and
$\psi_p = \braket{\hat{\sigma}_{Gp}} e^{i\omega_pt}$, the
semi-classical Heisenberg-Langevin equations of motion become
\begin{subequations}\label{eq:mean_field_1}
\begin{align}
  \partial_{t}\alpha_L &= ig_{\mathrm{eff}}^p \psi_p\alpha_S -
                         \kappa_{L}\alpha_L + i\sqrt{\kappa_L}\Phi_{\mathrm{in}},\\
  \partial_{t}\alpha_S &= ig_{\mathrm{eff}}^p \psi_p^{*}\alpha_L - \kappa_{S}\alpha_S,\\
  \partial_{t}\psi_p &= ig_{\mathrm{eff}}^p \alpha_S^{*}\alpha_L - \Gamma_p\psi_p.
\end{align}
\end{subequations}

The corresponding steady-state solutions (which agree with the
classical treatment of an OPO \cite{Yariv1966}) can be parametrized in
terms of $f = \Phi_{\mathrm{in}}/\Phi_{\mathrm{th}}$, where
$\Phi_{\mathrm{th}} = \sqrt{\kappa_{L}\kappa_{S}\Gamma_p} /
g_{\mathrm{eff}}^p$
is the threshold value for the driving parameter. Below threshold
($f<1$), neither the polariton nor the Stokes mode are populated
($|\psi_p|^2=|\alpha_S|^2=0$), while the pump mode has population
$|\alpha_L|^2 = f^2 \Phi_{\mathrm{th}}^2/\kappa_L$. Above threshold
($f\geq1$), the pump amplitude becomes independent of the driving
power (so-called pump clamping),
$|\alpha_L|^2=\Phi_{\mathrm{th}}^2/\kappa_L$, while the Stokes and
polariton mode occupations grow linearly with input power,
$|\psi_p|^{2} = (f-1) \Phi_{\mathrm{th}}^2/\Gamma_p$ and
$|\alpha_{S}|^{2} = (f-1) \Phi_{\mathrm{th}}^2 / \kappa_S$.
This implies that the conversion efficiency approaches $100\%$ if the
pumping is sufficiently strong. Explicitly, the quantum yield for
conversion of input photons to pairs of Stokes photons and polaritons
follows the simple relation
\begin{equation}
  \mathcal{Q} = \frac{P_{S}/\omega_{S}}{P_{\mathrm{in}}/\omega_{L}}
  = 1-\frac{1}{f},\label{eq:yield}
\end{equation}
where $P_{S}/\omega_{S}=\kappa_S|\alpha_S|^2$ ($=P_{p}/\omega_{p}$) is
the flux of emitted Stokes photons, and
$P_{\mathrm{in}}=\omega_L \Phi_{\mathrm{in}}\Phi_{\mathrm{th}}$
is the input power.

The number of photons emitted at the vibro-polariton frequency
(typically in the mid-IR~\cite{Shalabney2015}) is equal to the number
of generated Stokes photons, multiplied by the radiative emission
efficiency of the polaritons,
$\beta = \Gamma_p^{\mathrm{rad}}/\Gamma_p$. For zero detuning and a
mid-IR cavity without nonradiative losses (such as a dielectric
cavity~\cite{Muallem2016}), this is given by
$\beta = \frac{\kappa_c}{\kappa_c+\gamma_v}$, which is close to unity
for the experimentally relevant regime $\kappa_c\gg\gamma_v$. In a
standard Raman laser, the energy deposited into the vibrational modes
is converted to heat, limiting the achievable
powers~\cite{Pask2003,Boyd2008}. In contrast, the vibro-polariton
Raman OPO proposed here converts this energy efficiently into an
additional coherent output beam at mid-IR frequencies, and thus
simultaneously reduces heating significantly.

Furthermore, the ratio between the thresholds for polariton-based OPO
operation under strong coupling and for the bare-molecule Raman laser
under weak coupling is given by
\begin{equation}
  \frac{\Phi_{\mathrm{th}}^{SC}}{\Phi_{\mathrm{th}}^{WC}} 
  = \sqrt{\frac{\Gamma_p}{\gamma_v+\gamma_{\varphi}}} 
  \approx \sqrt{\frac{\kappa_c}{2\gamma_\varphi}}.
\end{equation}
This demonstrates that for the common case that the inhomogeneous
width and dephasing of the vibrational modes are faster than the
cavity losses ($\gamma_\varphi>\kappa_c$), the vibro-polariton Raman
OPO has a lower threshold power than the equivalent Raman laser. In
addition, depending on the relative lifetimes of the vibro-polaritons
$\Gamma_p$ and the Stokes photons $\kappa_S$, there can be significant
accumulation of population in the vibro-polariton mode, suggesting a
roadmap towards achieving vibro-polariton condensation (in analogy to
exciton-polariton condensation~\cite{Kasprzak2006}) based on the high
efficiency of SRS.

\begin{figure*}
  \includegraphics[width=\linewidth]{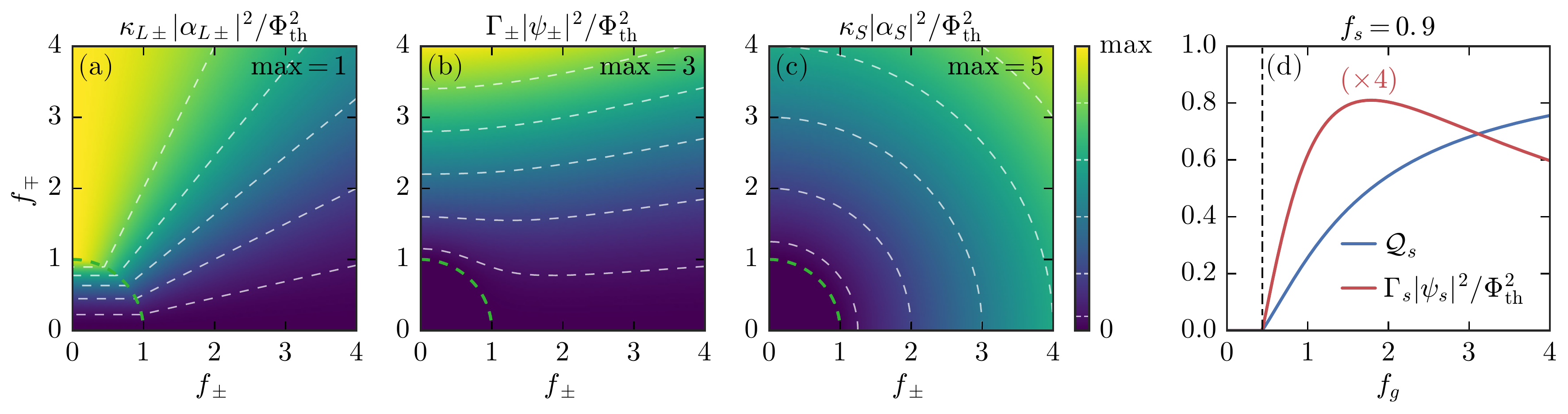}
  \caption{Rescaled population densities under two-mode pumping with
    $\Phi_{\mathrm{th}}^+=\Phi_{\mathrm{th}}^-$, for (a) the pump
    mode, (b) the polaritons and (c) the Stokes mode. The green dashed
    semicircles denote the threshold condition
    $f_{\mathrm{tot}}\geq1$. (d) Quantum efficiency $\mathcal{Q}_S$
    (blue) and rescaled signal polariton density (red, multiplied by
    $4$ for clarity) at a signal pump strength of $f_s=0.9$ as a
    function of the gate pump strength $f_g$.}
  \label{fig:populations_switchyield}
\end{figure*}

We next show how the coexistence of two vibro-polariton modes with
similar properties allows to turn the system into an all-optical
switch where emission at one frequency is switched by input at another
frequency~\cite{Gibbs1985,Dawes2005}. This is achieved by including a second
pump field, with the two pump frequencies chosen to make the Raman
process to the two polariton modes $\ket{+}$ and $\ket{-}$ resonant
with the same Stokes frequency,
\begin{equation}
  \omega_{L\pm}=\omega_{S}+\omega_{\pm},\label{eq:bimodal_cons}
\end{equation}
as depicted in \autoref{fig:setup}(b). Following the procedure of
adiabatic elimination and again performing a second RWA to remove
terms rotating at frequencies $\pm\Omega_R$ (see \cite{supplemental}
for details), we obtain the new effective Hamiltonian
\begin{multline}
  \hat{H}_{\mathrm{eff}}^{(2)} \simeq \omega_S\hat{n}_S + \sum_{\eta=\{\pm\}} \Big[ \omega_{L\eta}\hat{n}_{L\eta} + \omega_{\eta}\hat{\sigma}_{\eta\eta}\\
  - g^{\eta}_{\mathrm{eff}}
  \left(\hat{\sigma}_{G\eta}\hat{a}_S\hat{a}_{L\eta}^{\dagger} +
  \mathrm{H.c.}\right) \Big],
\end{multline}
with corresponding Heisenberg-Langevin equations in the mean-field
approximation
\begin{subequations}\label{eq:time_evo_2_pumps}
  \begin{align}
    \partial_{t}\alpha_{L\pm} &= ig^{\pm}_{\mathrm{eff}}\psi_{\pm}\alpha_{S}-\kappa_{L}\alpha_{L\pm}+i\sqrt{\kappa_{L\pm}}\Phi_{\mathrm{in}}^\pm,\\
    \partial_{t}\alpha_{S} &= ig^+_{\mathrm{eff}} \psi_+^{*} \alpha_{L+} + ig^-_{\mathrm{eff}} \psi_-^{*} \alpha_{L-} - \kappa_{S}\alpha_{S},\label{eq:common_Stokes}\\
    \partial_{t}\psi_{\pm} &= ig^{\pm}_{\mathrm{eff}} \alpha_{S}^{*} \alpha_{L\pm} - \Gamma_{\pm}\psi_{\pm}.
  \end{align}
\end{subequations}
The basic idea for achieving all-optical switching is then to use one
of the pump lasers as the input signal ($s = \pm$) and the other pump
laser as a gate ($g=\mp$). If the gate beam is turned off, the system
is identical to the OPO discussed up to now, and a weak signal beam
($f_s=\Phi_{\mathrm{in}}^s/\Phi_{\mathrm{th}}^s<1$) will not lead to
lasing, such that the corresponding polaritonic mode is not
populated. On the other hand, if the gate beam is strong enough to
support OPO operation ($f_g>1$), the Raman scattering for even a weak
signal beam is stimulated by the macroscopic population of the Stokes
mode, $|\alpha_S|^2\gg1$. We next demonstrate this idea in more detail
by solving for the steady state.

The relative phases of the different modes are fixed in the steady
state, leading to five equations only involving the absolute
amplitudes,
\begin{subequations}\label{eq:mods}
  \begin{align}
    \kappa_{S}|\alpha_{S}| &= \sum_{\eta=\{\pm\}}g_{\mathrm{eff}}^{\eta}|\psi_{\eta}||\alpha_{L\eta}|,\\	
    \Gamma_{\pm}|\psi_{\pm}| &= g^{\pm}_{\mathrm{eff}}|\alpha_{L\pm}||\alpha_{S}|,\\
    \kappa_{L\pm}|\alpha_{L\pm}| &= \sqrt{\kappa_{L\pm}}\Phi_{\mathrm{in}}^{\pm} - g^{\pm}_{\mathrm{eff}}|\psi_{\pm}||\alpha_{S}|.
\end{align}
\end{subequations}

These equations can be reduced to a quartic polynomial, which permits
an analytical solution. The general case is treated in the
supplemental material~\cite{supplemental}, while we here focus on the
case that the two thresholds are identical,
$\Phi_{\mathrm{th}}^+=\Phi_{\mathrm{th}}^-=\Phi_{\mathrm{th}}$, which
allows for simple analytical expressions. In particular, the threshold
condition can then be simplified to $f_{\mathrm{tot}}>1$, where
$f_{\mathrm{tot}} = \sqrt{f_+^2+f_-^2}$. Below threshold
($f_{\mathrm{tot}}<1$), the mean-field populations are identical to in
the single-pump case, with neither the polariton nor the Stokes modes
being populated ($|\psi_\pm|^2=|\alpha_S|^2=0$), while the pump mode
populations are just determined by the driving of each mode,
$|\alpha_{L\pm}|^2 = f_\pm^2 \Phi_{\mathrm{th}}^2/\kappa_L$. Above
threshold ($f_{\mathrm{tot}}\geq1$), the Stokes and polariton mode
occupations are given by
$|\alpha_{S}|^{2} = (f_{\mathrm{tot}}-1) \Phi_{\mathrm{th}}^2 /
\kappa_S$
and
$|\psi_\pm|^{2} = (f_{\mathrm{tot}}-1) \Phi_{\mathrm{th}}^2
f_\pm^2/(f_{\mathrm{tot}}^2\Gamma_\pm)$.
In contrast to the single-mode OPO case, the pump mode populations are
not clamped to a fixed value above threshold, but are given by
$|\alpha_{L\pm}|^{2} = \Phi_{\mathrm{th}}^2
f_{\pm}^2/(f_{\mathrm{tot}}^2\kappa_{L\pm})$.
The input power $P_{\mathrm{in}}^{\pm}$ in each pump mode thus does
not depend only on the external driving parameter
$\Phi^{\pm}_{\mathrm{in}}$, but also on the driving of the other mode
$\Phi^{\mp}_{\mathrm{in}}$. The mode populations as a function of
$f_+$ and $f_-$ are shown in \autoref{fig:populations_switchyield}. In
particular, it should be noted that there is only a single threshold,
below which no stimulated emission occurs, and above which all three
output modes are populated. Analysis of the fluctuations around the
steady-state values demonstrates that the obtained solutions are
stable~\cite{supplemental}. Thus, \emph{both} polariton modes show
stimulated emission due to the population of the Stokes mode as soon
as the total pump power becomes large enough. Consequently, the
quantum yield for conversion from each pump mode to the corresponding
polariton mode,
$\mathcal{Q}_{\pm} = \frac{P_{\pm}/\omega_{\pm}}
{P_{\mathrm{in}}^{\pm}/\omega_{L\pm}}$, becomes
\begin{equation}
  \mathcal{Q}_{+}=\mathcal{Q}_{-}=1-\frac{1}{f_{\mathrm{tot}}},
\end{equation}
where
$P_{\mathrm{in}}^{\pm} =
\omega_{L\pm}\sqrt{\kappa_{L\pm}}\Phi^{\pm}_{\mathrm{in}}|\alpha_{L\pm}|$
is the input power in pump mode $L\pm$~\cite{supplemental}. In
contrast to the ``normal'' OPO case in \autoref{eq:yield}, the quantum
yield of a given polariton does not depend on the corresponding input
power ($\propto f_{\pm}^2$), but only on the total one
($\propto f_{\mathrm{tot}}^2$). This demonstrates that the system can
indeed be used like a switch, as sketched above: A below-threshold
signal beam input $f_s<1$ does not produce output in the signal
polariton if the gate beam is turned off, but is efficiently converted
to signal polaritons if the gate is switched on ($f_g^2>1-f_s^2$). The
conversion efficiency of the signal can be made high by making the
gate beam sufficiently strong, as demonstrated in
\autoref{fig:populations_switchyield}(d). The switching speed is
limited by the lifetime of the longest-lived state in the system,
leading to a tradeoff between achieving low thresholds (requiring
small losses) and fast switching speeds (requring large losses).

To conclude, we have demonstrated that by taking advantage of the
phenomenon of collective vibrational strong coupling, it is feasible
to transform a Raman laser into an OPO. Apart from the improvement of
generating two coherent beams both in the visible and in the mid-IR
ranges, this new type of OPO presents a lower threshold and less heat
generation when compared to a standard Raman laser. Moreover, thanks
to the existence of two similar vibro-polaritons, this OPO can also
operate as an all-optical switch when excited by two properly designed
external beams. Our finding is thus an example of the great potential
that hybrid light-matter states possess in both manipulating light
fields and modifying material properties.

\acknowledgments
This work has been funded by the European Research Council
(ERC-2011-AdG proposal No. 290981), by the European Union Seventh
Framework Programme under grant agreement FP7-PEOPLE-2013-CIG-618229,
and the Spanish MINECO under contract MAT2014-53432-C5-5-R and the
``María de Maeztu'' programme for Units of Excellence in R\&D
(MDM-2014-0377).

\clearpage

\onecolumngrid
\begin{center}
\textbf{\large Supplemental material}
\end{center}
\vspace{\columnsep}
\twocolumngrid

\setcounter{equation}{0}
\setcounter{figure}{0}
\setcounter{table}{0}
\setcounter{page}{1}
\makeatletter
\renewcommand{\theequation}{S\arabic{equation}}
\renewcommand{\thefigure}{S\arabic{figure}}
\renewcommand{\bibnumfmt}[1]{[S#1]}
\renewcommand{\citenumfont}[1]{S#1}

\section{Derivation of the effective Hamiltonian}\label{sec:adiabatic}

In this section we illustrate the derivation of the OPO Hamiltonian,
as given by Eq.~(4) in the main text. We also sketch the extension to
the case of an all-optical switch with two input beams. We first
describe the adiabatic elimination of the electronically excited
states in detail. We assume that the detuning
$\Delta=\omega_e-\omega_L$ is large compared with the relevant energy
scales for the electronic ground states ($\ket{g^{(i)}}$,
$\ket{v^{(i)}}$), and that the interaction terms involving electronic
excitations in Eq.~(2) in the main text are perturbative. In the
following, we denote these terms as
$\hat{V} = \hat{V}_+ + \hat{V}_+^\dagger$, where $\hat{V}_+$ contains
all the terms \emph{creating} electronic excitations, The occupation
of the electronically excited states $\ket{e^{(i)}}$ with free
evolution Hamiltonian
$\hat{H}_e=\omega_e \sum_{i=1}^N\hat{\sigma}_{ee}^{(i)}$ is hence
vanishingly small. In particular, this assumption enables to avoid the
coupling of different electronically excited states, simplifying the
following treatment. In the adiabatic elimination procedure
\cite{S-Feshbach1958}, the density matrix equations are solved by
assuming a slow evolution of the lowest-lying states $\ket{g^{(i)}}$,
$\ket{v^{(i)}}$ and the optical modes $\hat{a}_S$, $\hat{a}_L$,
determined by
$\hat{H}_g = \hat{H}_s + \omega_S \hat{n}_S + \omega_{L} \hat{n}_L$,
with $\hat{H}_s$ given in Eq.~(1) in the main text. We here ignore
contributions originating from the incoherent dynamics within the
electronic excited manifold, which could be introduced by means of
effective Lindblad terms \cite{S-Reiter2012}, but are negligible for
large detuning. To perform the adiabatic approximation, we \emph{i)}
apply the \emph{rotating-frame} transformation
$\hat{U}=e^{-i(\omega_{S}\hat{n}_{S}+\omega_{L}\hat{n}_{L})t}$ and
\emph{ii)} we work in the eigenbasis of $\hat{H}_s$. The resulting
Hamiltonian is
\begin{equation}
  \hat{H}' = \hat{H}_{s} - \frac{1}{2} \hat{V}_{+}^{\prime\dagger}(t)
  \sum_{f,l}\frac{\omega_e}{\omega_e-\omega_f-\omega_l}\hat{H}_e^{-1}
  \hat{v}_+^{(l,f)}e^{i\omega_ft}.
  \label{S-eq:effective}
\end{equation}
Here, $\hat{V}'_+(t)$ has been expanded in terms of its frequency
components $f\in(L,S)$, as well as the system eigenstates
$l\in(+,-,\{d\})$ that it couples to, giving
\begin{multline}
  \hat{V}'_{+}(t) = \sum_{f,l}\hat{v}_+^{(f,l)}e^{i\omega_ft} = \sum_{i=1}^{N} \Bigg[
  g_{L}\hat{a}_{L}\hat{\sigma}_{Ge}^{\dagger(i)}e^{i\omega_{L}t} \\
  + g_{S}\hat{a}_{S} \left( \sum_{\eta=\{\pm\}} \frac{\hat{\sigma}_{\eta e}^{\dagger(i)}}{\sqrt{2N}}  +
    \sum_{d}u_{id}\hat{\sigma}_{de}^{\dagger(i)}\right)
  e^{i\omega_{S}t}\Bigg].
\label{S-eq:int_eig}
\end{multline} 
Here, $\hat{\sigma}_{le}^{(i)} = \ket{l}\bra{e^{i}}$, and the
coefficients appearing between parenthesis follow from the eigenstate
expansions
$\ket{v^{(i)}}=(2N)^{-\frac{1}{2}}\sum_{\eta=\{\pm\}}\ket{\eta}+\sum_d
u_{id}\ket{d}$,
where $u_{id}$ is the overlap matrix element between the $i$th
vibrational excitation and dark state $d$. These coefficients fulfill
$\sum_{i=1}^N u_{id}=0$ and are further constrained by the
orthogonality relation $\sum_{i=1}^N u_{di}u_{id'}=\delta_{dd'}$.
After going back to the nonrotating frame, the resulting effective
interaction reads
\begin{equation}\label{S-eq:int_OPO}
  \hat{H}_{\mathrm{int}}^{\mathrm{eff}} = -\sum_{\eta=\{\pm\}}
  g_{\mathrm{eff}}^{\eta}
  (\hat{a}_{L}\hat{a}_{S}^{\dagger}\hat{\sigma}_{G\eta}^{\dagger} + \mathrm{H.c.}),
\end{equation}
with
$g_{\mathrm{eff}}^{\eta} = \frac{g_{S}g_{L}}{2} \sqrt{\frac{N}{2}}
\left[(\omega_{e}-(\omega_{\eta}+\omega_{S}))^{-1} +
  \Delta^{-1}\right]$.
Note here that the contribution of the dark states is identically zero
in the effective dynamics. This is due to the fact that we assumed
perfect overlap between the involved modes, i.e., we took $g$, $g_S$,
and $g_L$ to be constant for all involved molecules. Relaxing this
condition would give an additional overlap prefactor
$g_{\mathrm{eff}}^{\eta}\to \mathcal{S}g_{\mathrm{eff}}^{\eta}$, with
$\mathcal{S}\propto \sum_i u_{i\eta} g_{S,i} g_{L,i}^*$, and also
give nonzero coupling to the dark states, but would not otherwise
change the results presented in the main text. In addition to the
effective interaction, we obtain (nonlinear) energy shifts, given by
\begin{equation}\label{S-eq:e_shifts}
  \hat{H}_{\mathrm{shift}}^{\mathrm{eff}} = 
  -\frac{g_L^2N}{\Delta}\hat{n}_L\hat{\sigma}_{GG}
  -\sum_l \frac{g_{S}^{2} \hat{n}_S \hat{\sigma}_{ll}} {\omega_{e}-(\omega_{l}+\omega_{S})}.
\end{equation}
Under the assumption that the output modes are not significantly
populated ($\hat{\sigma}_{GG}\approx1$, $\hat{n}_S,\hat{\sigma}_{\eta\eta}\ll1$), the
first term just gives a constant energy shift (which we assume to be
included into $\omega_L$), while the second term can be
neglected. This is a good approximation for typical system parameters
even when a large number of output photons is generated, due to the
relatively short lifetime of the polaritons. Under this approximation,
the vibro-polaritons are well-modeled as bosons
$[\hat{\sigma}_{G\eta},\hat{\sigma}_{G\eta'}^{\dagger}]\simeq\delta_{\eta\eta'}$,
and the trilinear interaction \autoref{S-eq:OPO_eff} corresponds to a
nondegenerate OPO under the identifications $\hat{a}_S\rightarrow$
signal, $\hat{\sigma}_{Gp}\rightarrow$ idler.

For a laser frequency chosen such that Raman scattering to one of the
polaritonic modes is resonant with the cavity mode $S$ in the optical,
$\omega_L=\omega_S+\omega_p$, with $p\in\{\pm\}$, there are rapidly
oscillating terms in \autoref{S-eq:int_OPO} in the interaction picture
with regards to $\hat{H}_g$. Averaging over a time sufficiently big
compared to $\tau_{\mathrm{coh}}\sim\Omega_R^{-1}$, these
contributions, which correspond to the coupling of the laser field
with the detuned polariton, can be neglected under a \emph{second}
rotating wave approximation, giving
\begin{align}
\hat{H}_{\mathrm{eff}}\simeq\omega_L\hat{n}_L+\omega_S\hat{n}_S+\omega_{p}\hat{\sigma}_{pp}-g_{\mathrm{eff}}^{p}\left(\hat{a}_{L}\hat{a}_{S}^{\dagger}\hat{\sigma}_{Gp}^{\dagger}+\mathrm{H.c.}\right).\label{S-eq:OPO_eff}
\end{align}

We next sketch the derivation of the effective Hamiltonian under
pumping of multiple input modes, as given by Eq.~(11) in the main
text. The pumping Hamiltonian is then
\begin{equation}
  \hat{H}^{(2)}_{d} = \sum_{\eta=\{\pm\}} \sqrt{\kappa_{L\eta}}\Phi_{\mathrm{in}}^{\eta}(\hat{a}_{L\eta}e^{-i\omega_{L\eta}t}+\hat{a}_{L\eta}^{\dagger}e^{i\omega_{L\eta}t}),
\end{equation}
while the pump-system interaction is given by
\begin{equation}
  \hat{V}_+ =
  \sum_{i=1}^{N}\big(g_{S}\hat{a}_S\hat{\sigma}_{v_ie_i}^{\dagger} +
  \sum_{\eta=\{\pm\}}g_{L\eta}\hat{a}_{L\eta}
  \hat{\sigma}_{g_ie_i}^{\dagger}\big).
\end{equation}
The frequencies $\omega_{L\pm}$ are chosen to satisfy the resonance
conditions for both polaritons,
$\omega_{L\pm}=\omega_{S}+\omega_{\pm}$. The adiabatic elimination of
the electronic states proceeds analogously to the single-pump case,
giving the effective interaction Hamiltonian
\begin{equation}\label{S-eq:int_OPO_2}
  \hat{H}_{\mathrm{int}}^{\mathrm{eff}(2)} =  -\sum_{\eta,\eta'=\{\pm\}}g_{\mathrm{eff}}^{\eta,\eta'}
  (\hat{a}_{L\eta'}\hat{a}_{S}^{\dagger}\hat{\sigma}_{G\eta}^{\dagger} + \mathrm{H.c.}),
\end{equation}
where the effective coupling constant of the pump field $\eta$ with
the polariton $\eta'$ is (using $\Delta_\pm=\omega_e-\omega_{L\pm}$)
\begin{equation}
  g_{\mathrm{eff}}^{\eta,\eta'} = \frac{g_{S}g_{L\eta}}{2} \sqrt{\frac{N}{2}} 
  \left[\frac{1}{\omega_{e}-(\omega_{\eta'}+\omega_{S})}+\frac{1}{\Delta_\eta}\right].
\end{equation}
The off-diagonal terms $\eta\not=\eta'$ can be neglected under the
same \emph{second} RWA we invoked for the OPO case. In addition to the
effective interaction, we again obtain extra nonlinear terms, given by
\begin{multline}\label{S-eq:H2_extra}
  \hat{H}_{\mathrm{extra}}^{\mathrm{eff}(2)} = -\sum_{\eta,\eta'} \Lambda_{\mathrm{eff}}^{\eta,\eta'} \hat{\sigma}_{GG} (\hat{a}_{L\eta}^{\dagger}\hat{a}_{L\eta'} + \mathrm{H.c.})\\
  - \hat{n}_S\sum_{l} \frac{g_{S}^{2}\hat{\sigma}_{ll}}{\omega_{e}-(\omega_{l}+\omega_{S})},
\end{multline}
which in addition to the energy shifts already seen in the single-pump
OPO case also contains an extra crossed term coupling the two pump
fields. The nonlinear terms \autoref{S-eq:H2_extra} can again be
neglected under the low-occupation assumption and the second
RWA. Finally, we thus obtain
\begin{multline}
  \hat{H}^{(2)}_{\mathrm{eff}}\simeq \omega_S \hat{n}_S +
  \sum_{\eta=\{\pm\}} \bigg[ \omega_{L\eta}\hat{n}_{L\eta} + \omega_{\eta}\hat{\sigma}_{\eta\eta}\\
  - g^{\eta}_{\mathrm{eff}}
  (\hat{\sigma}_{G\eta}\hat{a}_S\hat{a}_{L\eta}^{\dagger} +
  \mathrm{H.c.})\bigg],
  \label{S-eq:H_2}
\end{multline}
where we used that $g_{\mathrm{eff}}^{\eta,\eta} = g_{\mathrm{eff}}^{\eta}$.

\section{Mean-field steady-state solutions}
We here discuss the general steady-state solution for the
optical-switch setup with multiple pump beams. For clarity, we first
recall the well-known Manley-Rowe relations for beam fluxes and
powers~\cite{S-Yariv1966} in the single-pump case, which follow
straightforwardly from the semiclassical equations Eq.~(7) in the main
text. They connect the fluxes of emitted photons
$P_i/\omega_i=\gamma_i|\alpha_i|^2$ in the different modes, with the
simple relation
\begin{align}
  \frac{P_S}{\omega_S}&=\frac{P_p}{\omega_p}=\frac{P_L^{\mathrm{th}}}{\omega_L}(f-1),
  & f&=\frac{\Phi_{\mathrm{in}}}{\Phi_{\mathrm{th}}}. \label{S-eq:conservation}
\end{align}
This explicitly shows that Raman scattering converts each of the
incoming pump photons into a Stokes photon/polariton pair. From
\autoref{S-eq:conservation} and the resonance condition
$\omega_L=\omega_S+\omega_p$, we obtain the power relation
$P_{\mathrm{in}}=P_S +P_p+P_L^{\mathrm{th}}$. This expresses the fact
that the input power
$P_{\mathrm{in}}= \omega_L\Phi_{\mathrm{in}}\Phi_{\mathrm{th}}$ is
shared among the three modes, with a maximum clamped power for the $L$
mode at threshold equal to $\Phi_{\mathrm{th}}^2$. From the analogue
steady-state relations within the two-pump scenario, a set of
generalized Manley-Rowe relations accounting for the exchange of
energy between the modes participating in the scattering holds:
\begin{align}\label{S-eq:power_rels_2}
  \frac{P_{S}}{\omega_{S}} &= \sum_{\eta=\{\pm\}} \frac{P_{\eta}}{\omega_{\eta}},
  &\frac{P_{\pm}}{\omega_{\pm}} &= \frac{P_{\mathrm{in}}^{\pm}-P_{L\pm}}{\omega_{L\pm}},
\end{align}
where the input power in each of the pump modes $L\pm$ is
$P_{\mathrm{in}}^{\pm}=\omega_{L\pm}\sqrt{\kappa_{L\pm}}\Phi^{\pm}_{\mathrm{in}}|\alpha_{L\pm}|$.
Employing the resonance conditions for the two pumps, we obtain the
global power relation
$\sum_{\eta=\{\pm\}}P_{\mathrm{in}}^\eta = P_S + \sum_{\eta=\{\pm\}}
\left(P_\eta+P_{L\eta}\right)$,
a direct generalization of \autoref{S-eq:conservation}.

We now proceed to solve the steady-state equations, Eq.~(13) in the
main text, under pumping of both modes ($\Phi_\mathrm{in}^\pm>0$). The
following relations between the $L\pm$ amplitudes hold,
\begin{equation}
  \frac{\kappa_{L+}|\alpha_{L+}|^{2}}{\Phi_{\mathrm{th}}^{+2}}+\frac{\kappa_{L-}|\alpha_{L-}|^{2}}{\Phi_{\mathrm{th}}^{-2}}=1,\label{S-eq:general_eqs}
\end{equation}
expressing the fact that the \emph{global} pump amplitude becomes
clamped above the threshold due to its connection to a common Stokes
mode (with
$\Phi_{\mathrm{th}}^{\pm} =
\sqrt{\kappa_{L\pm}\kappa_{S}\Gamma_{\pm}}/g^\pm_{\mathrm{eff}}$).
As we will see in the following, this is not the case for each of the
pumping amplitudes individually. The relation \autoref{S-eq:general_eqs}
suggests we can define
\begin{align}
|\alpha_{L+}|=\frac{\Phi_{\mathrm{th}}^{+}}{\sqrt{\kappa_{L+}}}\sin\Theta,&& |\alpha_{L-}|=\frac{\Phi_{\mathrm{th}}^{-}}{\sqrt{\kappa_{L-}}}\cos\Theta,
\end{align}
with mixing angle $\Theta$ in the range $\Theta\in(0,\pi/2)$, such
that $|\alpha_{L\pm}|>0$. Inserting this into the steady
state-equations leads to a quartic equation for $t=\tan(\Theta/2)$,
given by
\begin{equation}\label{S-eq:quartic}
  t^{4}+2(\alpha-\beta)t^{3}+2(\alpha+\beta)t-1=0,
\end{equation}
where
$\alpha=\Phi_{\mathrm{in}}^{-}\Phi_{\mathrm{th}}^{-}/(\Phi_{\mathrm{in}}^{+}\Phi_{\mathrm{th}}^{+})$,
and
$\beta=[(\Phi_{\mathrm{th}}^{+})^{2}-(\Phi_{\mathrm{th}}^{-})^{2}]/(\Phi_{\mathrm{in}}^{+}\Phi_{\mathrm{th}}^{+})$.
We have checked that this equation has only one physical solution
$0<t<1$ for arbitrary values of $\alpha>0$) and $\beta$. The
analytical form of the solution of \autoref{S-eq:quartic} is very
lengthy and we thus omit it in the following. However, in the
degenerate case with equal thresholds, we get $\beta=0$ and the
equation can be factorized as
$\left(t^2+1\right) \left(t^2+2 \alpha t-1\right)=0$, leading to the
single physical solution analyzed in the main text,
$t=\sqrt{1+\alpha ^2}-\alpha$, with $\alpha=f_-/f_+$.

\section{Stability of the mean-field solutions}
\allowdisplaybreaks

We here analyze the stability of the semiclassical steady-state
solutions of Eq.~(12) in the main text. Collecting these solutions in
the vector
$\boldsymbol{v}^{\infty}=(\alpha_S,\psi_+,\alpha_{L+},\psi_-,\alpha_{L-},\mathrm{H.c.})$,
inserting the linearized solution
$\boldsymbol{v}(t)=\boldsymbol{v}^{\infty}+\delta\boldsymbol{v}(t)$ in
Eq.~(12) and keeping terms $O(\delta\boldsymbol{v})$, we obtain the
time evolution of the fluctuations,
$\partial_{t}\delta\boldsymbol{v}(t) =
\boldsymbol{\mathcal{M}}\delta\boldsymbol{v}(t)$.
The stability matrix is
\begin{equation}
  \boldsymbol{\mathcal{M}} = \left(\begin{array}{c|cc}
    -\kappa_{S}\mathbb{1}_{2\times2} & \boldsymbol{v}_{+} & \boldsymbol{v}_{-}\\
    \hline \boldsymbol{u}_{+}^T & -\boldsymbol{\mathcal{P}}_{+} & 0\\
    \boldsymbol{u}_{-}^T & 0 & -\boldsymbol{\mathcal{P}}_{-}
  \end{array}\right),
\end{equation}
where the submatrices are
\begin{subequations}
  \begin{equation}
    \boldsymbol{v}_\pm = ig_{\mathrm{eff}}^{\pm} \begin{pmatrix}
      0 & \alpha_{L\pm} & \psi_\pm^{*} & 0\\
      -\alpha_{L\pm}^{*} & 0 & 0 & -\psi_\pm
    \end{pmatrix},
  \end{equation}
  \begin{align}
    \boldsymbol{u}_\pm &= ig_{\mathrm{eff}}^{\pm} \begin{pmatrix}
      \alpha_{L\pm} & 0 & 0 & -\psi_\pm^{*}\\
      0 & -\alpha_{L\pm}^{*} & \psi_\pm & 0
    \end{pmatrix},\\
    \boldsymbol{\mathcal{P}}_{\pm} &= \begin{pmatrix}
      \Gamma_{\pm} & 0 & -ig_{\mathrm{eff}}^{\pm}\alpha_{S}^{*} & 0\\
      0 & \Gamma_{\pm} & 0 & ig_{\mathrm{eff}}^{\pm}\alpha_{S}\\
      -ig_{\mathrm{eff}}^{\pm}\alpha_{S} & 0 & \kappa_{L\pm} & 0\\
      0 & ig_{\mathrm{eff}}^{\pm}\alpha_{S}^{*} & 0 & \kappa_{L\pm}
    \end{pmatrix}.
  \end{align}
\end{subequations}
The fluctuations $\delta\boldsymbol{v}(t)$ will grow exponentially in
time if the real part of any eigenvalue of $\boldsymbol{\mathcal{M}}$
is positive. This can be tested via the Routh-Huwirtz criterion
\cite{S-Gradshteyn2014}, which provides necessary and sufficient
conditions for the roots of the characteristic polynomial
$\det\big({\boldsymbol{\mathcal{M}}-\lambda\mathbb{1}_{10\times10}}\big)$
to have negative real part, without explicit knowledge of their
values. Applying this criterion proves that the all-optical switch
solutions pictured in the main text are stable. From this, the
stability of the OPO solutions under single-mode driving follows
automatically.

\end{document}